\documentclass[aps,prd,10pt,nofootinbib,twocolumn,eqsecnum,showpacs,showkeys,superscriptaddress,preprintnumbers,altaffilletter]{revtex4-1}

\usepackage{amssymb,amsmath}
\usepackage[utf8]{inputenc}
\usepackage{graphicx}
\usepackage{url}
\usepackage{hyperref}
\usepackage{xcolor}
\usepackage{soul}

\def\be{\begin{equation}}
\def\ee{\end{equation}}

\def\bea{\begin{eqnarray}}
\def\eea{\end{eqnarray}}

\begin{document}


\title{Observational constraints of $f(Q)$ gravity}

\author{Ruth Lazkoz}
\email{ruth.lazkoz@ehu.es}
\affiliation{Department of Theoretical Physics, University of the Basque Country UPV/EHU, P.O. Box 644, 48080 Bilbao, Spain}
\author{Francisco S. N. Lobo}
\email{fslobo@fc.ul.pt}
\affiliation{Instituto de Astrofísica e Ci\^encias do EspaÇo, Facultade de Ci\^encias da Universidade de Lisboa, Edif\'icio C8, Campo Grande, P-1749-016, Lisbon, Portugal}
\author{Mar\'ia Ortiz-Ba\~nos}
\email{maria.ortiz@ehu.eus}
\affiliation{Department of Theoretical Physics, University of the Basque Country UPV/EHU, P.O. Box 644, 48080 Bilbao, Spain}
\author{Vincenzo Salzano}
\email{vincenzo.salzano@usz.edu.pl}
\affiliation{Institute of Physics, Faculty of Mathematics and Physics, University of Szczecin, Wielkopolska 15, 70-451 Szczecin, Poland}

\date{\today}

\begin{abstract}

In this work, we consider an extension of symmetric teleparallel gravity, namely, $f(Q)$ gravity, where the fundamental block to describe spacetime is the nonmetricity, $Q$. Within this formulation of gravitation, we perform an observational analysis of several modified $f(Q)$ models using the redshift approach, where the $f(Q)$ Lagrangian is reformulated as an explicit function of the redshift, $f(z)$. Various different polynomial parameterizations of $f(z)$ are proposed, including new terms which would allow for deviations from the $\Lambda$CDM model. Given a variety of observational probes, such as the expansion rate data from early-type galaxies, Type Ia Supernovae, Quasars, Gamma Ray Bursts, Baryon Acoustic Oscillations data and Cosmic Microwave Background distance priors, we have checked the validity of these models at the background level in order to verify if this new formalism provides us with plausible alternative models to explain the late time acceleration of the universe. Indeed, this novel approach provides a different perspective on the formulation of observationally reliable alternative models of gravity.

\end{abstract}

\maketitle

\section{Introduction}

General Relativity is traditionally described in terms of the Levi-Civita connection, which conforms the basis of Riemannian geometry. This choice relies on the assumption of a torsion and nonmetricity free geometry. Within this framework, the Ricci curvature scalar $R$ acts as the building block of the spacetime. Although this is usually done for historical reasons it is important to keep in mind that the connection has a more general expression \cite{Hehl:1994ue, ortin} and General Relativity can be described in terms of different geometries from the Riemannian one. One of the alternatives is what is called teleparallel gravity \cite{book}, where the gravitational force is driven by the torsion, $T$. Although this formalism was formally proposed in \cite{book}, Einstein himself already used such a geometry in one of his unified field theory attempts \cite{eins}.
Another possible alternative is symmetric teleparallel gravity \cite{Nester:1998mp}, where one considers a vanishing curvature and torsion,
and it is the nonmetricity, $Q$, that mediates the gravitational interaction. Some other interesting cases can be thought as, for example, a geometry where torsion carries part of the gravitational force and nonmetricity carries the rest. Along with the increasing interest on extended theories of gravity, all these alternative geometries have been explored due to the fact that the intrinsic implications and features of the gravitational theories could be different to the ones corresponding to Riemannian geometry \cite{Ferraro:2008ey, Linder:2010py, Cai:2015emx, Geng:2011aj, Jarv:2015odu, BeltranJimenez:2017tkd, Conroy:2017yln}.

Indeed, by making assumptions on the affine connection, one is essentially specifying a metric-affine geometry \cite{Jarv:2018bgs}. Recall that the metric tensor $g_{\mu\nu}$ can be considered as a generalization of the gravitational potential, and it is essentially used to define notions such as angles, distances and volumes, while the affine connection $\Gamma^{\mu}{}_{\alpha\beta}$ defines parallel transport and covariant derivatives. In this context, as mentioned above, a basic result in differential geometry states that the general affine connection may be decomposed into the following three independent components \cite{Hehl:1994ue,ortin}:
\begin{equation}
	\label{Connection decomposition}
	\Gamma^{\lambda}{}_{\mu\nu} =
	\left\lbrace {}^{\lambda}_{\phantom{\alpha}\mu\nu} \right\rbrace +
	K^{\lambda}{}_{\mu\nu}+
	L^{\lambda}{}_{\mu\nu} \,,
\end{equation}
where $\left\lbrace {}^{\lambda}_{\phantom{\alpha}\mu\nu} \right\rbrace \equiv \frac{1}{2} g^{\lambda \beta} \left( \partial_{\mu} g_{\beta\nu} + \partial_{\nu} g_{\beta\mu} - \partial_{\beta} g_{\mu\nu} \right)$ is the Levi-Civita connection of the metric $g_{\mu\nu}$; the term $K^{\lambda}{}_{\mu\nu} \equiv \frac{1}{2} T^{\lambda}{}_{\mu\nu}+T_{(\mu}{}^{\lambda}{}_{\nu)} $ is the contortion, with the torsion tensor defined as $T^{\lambda}{}_{\mu\nu}\equiv 2 \Gamma^{\lambda}{}_{[\mu\nu]}  $; and finally the disformation $L^{\lambda}{}_{\mu\nu}$ is given by
\begin{equation}
	\label{Disformation}
	L^{\lambda}{}_{\mu\nu} \equiv \frac{1}{2} g^{\lambda \beta} \left( -Q_{\mu \beta\nu}-Q_{\nu \beta\mu}+Q_{\beta \mu \nu} \right)  \,,
\end{equation}
which is defined in terms of the nonmetricity tensor, $Q_{\alpha \mu \nu} \equiv \nabla_{\alpha} g_{\mu\nu}$.

In this work we will focus on a torsion and curvature free geometry, which is  exclusively defined by the nonmetricity $Q_{\alpha \mu \nu}$. As this is a novel approach, no cosmological tests have been carried out so far, and its exploration will hopefully offer some insight on the late accelerated expansion of the universe. Within the scenario of modified gravity alternatives, we will take as an initial point the idea of generalizing $Q$-gravity in an analogous manner to what has been done with $f(R)$ theories. We will start from the $f(Q)$-type of theories presented in \cite{BeltranJimenez:2017tkd} and rewrite some proposals in the redshift approach \cite{Lazkoz:2018aqk} in order to explore the cosmological background evolution within this kind of geometry.

More specifically, in \citep{Lazkoz:2018aqk}, the $f(R)$ Lagrangian was reformulated as an explicit function of the redshift, i.e., as $f(z)$. Thus,  various different polynomial parameterizations $f(z)$ were proposed, including new terms which would allow for deviations from the $\Lambda$CDM model, and were thoroughly confronted with observational data. In fact, this novel approach provides a different perspective for the development of new and observationally reliable models of gravity.

The work is organized as follows: We set the stage in Sec.~\ref{secII}, by briefly outlining the general formalism of $f(Q)$ gravity. In Sec. \ref{secIII}, we reformulate the $f(R)$ Lagrangian as an explicit function of the redshift, and provide reasonable $f(z)$ models from the outset, which are to be tested at a later stage, by numerically solving the gravitational field equations, in order to test their validity and study their deviation with respect to the $\Lambda$CDM scenario. In Sec.~\ref{sec:data} we describe the observational data used in our analysis, namely, the expansion rate data from early-type galaxies, Type Ia Supernovae, Quasars, Gamma Ray Bursts, Baryon Acoustic Oscillations data and Cosmic Microwave Background distance priors. In Sec.~\ref{sec:results}, we discuss our results and finally conclude in Sec. \ref{sec:conclusions}.

\section{Setting the stage: $f(Q)$ gravity}
\label{secII}

Consider the proposal of $f(Q)$ gravity given by the following action \cite{BeltranJimenez:2017tkd}:
\be  \label{qqm}
 S=\int \left[\frac{1}{2}f(Q)+\mathcal{L}_m\right]\sqrt{-g}~d^4x,
 \ee
where $f(Q)$ is an arbitrary function of the nonmetricity $Q$, $g$ is the determinant of the metric $g_{\mu\nu}$ and ${\cal L}_m$ is the matter lagrangian density.

The nonmetricity tensor is defined as
\begin{equation}
Q_{\alpha \mu \nu }=\nabla _{\alpha }g_{\mu \nu }\,,
\end{equation}%
and its two traces as follows:
\begin{equation}
Q_{\alpha}=Q_{\alpha }{}^{\mu }{}_{\mu }\,,\quad \tilde{Q}_{\alpha }=Q^{\mu
}{}_{\alpha \mu }\,.
\end{equation}%
It is also useful to introduce the superpotential
\begin{eqnarray}
4P^{\alpha }{}_{\mu \nu } &=& -Q^{\alpha }{}_{\mu \nu } + 2Q_{(\mu %
\phantom{\alpha}\nu )}^{\phantom{\mu}\alpha } - Q^{\alpha }g_{\mu \nu }  \notag
\\
&&-\tilde{Q}^{\alpha }g_{\mu \nu }-\delta _{(\mu }^{\alpha }Q_{\nu )}\,. \label{super}
\end{eqnarray}
One can readily check that $Q=-Q_{\alpha \mu \nu }P^{\alpha \mu \nu
}$ (with our sign conventions that are the same as in Ref. \cite{BeltranJimenez:2017tkd}).

The energy-momentum tensor is given by
\begin{eqnarray}
T_{\mu \nu } &=&-\frac{2}{\sqrt{-g}}\frac{\delta \sqrt{-g}\,{{\cal L}}_{m}}{\delta
g^{\mu \nu }}\,, \label{emt}
\end{eqnarray}
and for notational simplicity, we introduce the following definition
\begin{equation}
f_Q=f^{\prime }(Q)\,.  \label{f_F}
\end{equation}%

Varying the action (\ref{qqm}) with respect to the metric, one obtains the
gravitational field equation given by
\begin{eqnarray}
&&\frac{2}{\sqrt{-g}}\nabla_\alpha\left(\sqrt{-g} f_Q P^\alpha{}_{\mu\nu}\right)
+ \frac{1}{2}g_{\mu\nu} f  \notag \\
&& + f_Q \left( P_{\mu\alpha\beta}Q_{\nu}{}^{\alpha\beta}
-2Q_{\alpha\beta\mu}P^{\alpha\beta}{}_\nu\right) = - T_{\mu\nu} \,,\label{efe}
\end{eqnarray}
and varying (\ref{qqm}) with respect to the connection, one obtains
\begin{equation}
\nabla_\mu\nabla_\nu \left(\sqrt{-g} f_Q P^{\mu\nu}{}_\alpha \right) = 0\,.  \label{cfe}
\end{equation}

With the formalism of $f(Q)$ gravity specified, we will next consider cosmological applications, by reformulating the $f(Q)$ Lagrangian as an explicit function of the redshift, namely, as $f(z)$. Furthermore, by providing reasonable $f(z)$ models from the outset, i.e. at the level of the action, so that at a later stage one can numerically solve the dynamics of the universe, to test their validity and study their deviation with respect to the $\Lambda$CDM scenario.

\section{The $f(z)$ approach}\label{secIII}

\subsection{Cosmology}

Considering a FLRW universe represented by the following isotropic, homogeneous and spatially flat line element
\begin{equation}  \label{frw}
{\mathrm{d}} s^2 = -{\mathrm{d}} t^2 + a^2(t)\delta_{ij} {\mathrm{d}}
x^i {\mathrm{d}} x^j\,,
\end{equation}
and taking into account the energy-momentum tensor of a perfect fluid, given by $T_{\mu \nu}=\left(\rho+p\right)u_{\mu}u_{\nu}+pg_{\mu \nu}$, where $\rho $ and $p$ are the thermodynamic energy density and isotropic pressure, we obtain the Friedmann and Raychadhuri equations, given by \cite{Harko:2018gxr}
\bea\label{fried}
3H^2&=&\frac{1}{2 f_Q}\left(-\rho+\frac{f}{2}\right) \,,
	\\\label{ray}
\dot{H}+3H^2+\frac{\dot{f}_Q}{f_Q}H&=&\frac{1}{2f_Q}\left(p+\frac{f}{2}\right)  \,,
\eea
respectively, where the overdot is defined as $\cdot\equiv d/dt$. In addition to this, consider the energy conservation equation for standard energy-matter perfect fluids:
\be
\dot{\rho}_i+3H\left(\rho_i+3 p_i \right)=0,
\ee
where the suffix $i$ stands for matter, radiation or any other fluid in the stress-energy tensor.

As our goal is to propose $f(z)$ models and find the evolution of the background  we will manipulate   Eq. (\ref{fried}) and write it in terms of the redshift. We perform thus a change of variable taking into account that in a FLRW geometry, $Q=6H^2$ holds and then $f_Q=\frac{f_z}{6 H^2_z}$ getting
\be
H^2=\frac{(H^2)_z}{f_z}\left(-\rho+\frac{f}{2}\right).\label{f}
\ee
where the subindex $z$ denotes the derivative with respect to the redshift.

In this work we will follow an analogous approach to the one outlined in \cite{Lazkoz:2018aqk}: we will reformulate an $f(Q)$ model as an explicit function of redshift, $f(z)$, solve Eq.~(\ref{f}) numerically, and then we will apply the obtained $H(z)$ to observational data.

\subsection{Specific $f(z)$ proposals}

In order to select some $f(z)$ models that could be interesting to study we first take a look at an $f(Q)$ model that mimics a $\Lambda$CDM background expansion, which is $f_{\Lambda}(Q)=-Q$. This can be clearly seen replacing this expression in Eq. (\ref{f}). Assuming a universe filled with matter, radiation and a cosmological constant, the Friedmann equation reads
\be
H^2(z)=\Omega_m(1+z)^3+\Omega_r(1+z)^4+(1-\Omega_m-\Omega_r) \,,
\ee
where $\Omega_i=8\pi \rho_i/(3H_0^2)$ is the dimensionless density parameter, and $i=m,\,r$ refers to matter and radiation, respectively.

Then, $f_{\Lambda}(Q)$ in terms of $z$ gives
\be\label{fl}
f_{\Lambda}(z)=-6(1-\Omega_m-\Omega_r)-6\Omega_m(1+z)^3-6\Omega_r(1+z)^4.
\ee
Here we can notice that a $\Lambda$CDM background in the redshift formalism would be described by a constant term, a third order term and a fourth order one. Now, we would like to choose $f(z)$ models which are somehow generalizations of Eq. (\ref{fl}). To do that we have found useful to make use of the tendencies we observe in $f_{\Lambda}$. Even if $\Lambda$CDM is not the definitive model to explain the late-time expansion of the universe, it does describe this late phase quite satisfactorily. Then, it is reasonable to propose models which satisfy the trends it shows at low and high redshift, that is,
\bea
\lim_{z\rightarrow 0} f(Q(z))& \propto & {\rm const} \,,\\
\lim_{z\rightarrow \infty} f(Q(z))& \propto & (1+z)^4 \,.
\eea
In fact, due to the relation between $Q$ and $H$ this will be satisfied at least for all the dark energy models studied in \cite{Lazkoz:2018aqk}.

\subsection{Models considered}

As in \cite{Lazkoz:2018aqk}, we will propose simple but well-motivated polynomical generalizations of the most fundamental model $f_{\Lambda}(z)$. As the high and low redshift limits lie within a constant term and an order four polynomial, we have constructed our proposals by adding some terms consisting on powers of $(1+z)$ which are within these two limits. Moreover, we have selected small powers as this allows us to introduce small modifications that may induce changes in the background evolution but preserving the desired behaviour.

In the following analysis, we consider the following models:
\begin{enumerate}
\item $f_0+f_3 (1+z)^3+f_4 (1+z)^4$,
 \item $f_{12}(1+z)^{1/2}+f_3(1+z)^3+f_4 (1+z)^4$,
\item $f_{12}(1+z)^{1/2}+f_1(1+z)+f_2(1+z)^2+f_3(1+z)^3+f_ 4(1+z)^4$,
 \item $f_{14}(1+z)^{1/4}+f_3 (1+z)^3+f_4 (1+z)^4$,
 \item $f_{14}+f_1(1+z)+f_2(1+z)^2+f_3(1+z)^3+f_4 (1+z)^4$,
 \item $f_{16}(1+z)^{1/6}+f_3(1+z)^3+f_4 (1+z)^4$,
  \item $f_{16}(1+z)^{1/6}+f_1(1+z)+f_2(1+z)^2+f_3(1+z)^3+f_4 (1+z)^4$,
  \end{enumerate}
where the factors $f_i$ are free parameters corresponding to each model.

\section{Observational Data}\label{sec:data}

In this section, we outline the cosmological data used in this work. We use various current observational data to constrain the $f(z)$ models described in the previous section, and consider the data which is related to the expansion history of the universe, i.e., the ones describing the distance-redshift relations. More specifically, we use the expansion rate data from early-type galaxies, Type Ia Supernovae, Quasars, Gamma Ray Bursts, Baryon Acoustic Oscillations data and Cosmic Microwave Background distance priors.

\subsection{Hubble data}

Early-type galaxies (ETGs) displaying a passive  evolution provide Hubble parameter measurements through estimations of their differential evolution.
Compilations of such observations can be  regarded as cosmic chronometers, and we use a sample covering  the redshift range $0<z<1.97$, which received a recent update in \cite{10.1093/mnrasl/slv037}.
For these measurement one can construct a $\chi^2_{H}$ estimator as follows:
\begin{equation}\label{eq:hubble_data}
\chi^2_{H}= \sum_{i=1}^{24} \frac{\left( H(z_{i},\boldsymbol{\theta})-H_{obs}(z_{i}) \right)^{2}}{\sigma^2_{H}(z_{i})} \; ,
\end{equation}
where $\sigma_{H}(z_{i})$ stand for the observational errors on the measured values $H_{obs}(z_{i})$, and $\boldsymbol{\theta}$ is the vector of the cosmological background parameters.

\subsection{Pantheon Supernovae data}

One of the latest Type Ia Supernovae (SNeIa) data compilation is the Pantheon compilation \cite{Scolnic:2017caz}. We choose this  set of $1048$ SNe, which covers
 the redshift range $0.01<z<2.26$, and use it in the usual manner to define
\begin{equation}
\chi^2_{SN} = \Delta \boldsymbol{\mathcal{F}}^{SN} \; \cdot \; \mathbf{C}^{-1}_{SN} \; \cdot \; \Delta  \boldsymbol{\mathcal{F}}^{SN} \; .
\end{equation}
Here $\Delta\boldsymbol{\mathcal{F}} = \mathcal{F}_{\rm theo} - \mathcal{F}_{\rm obs}$ represents the difference between the theoretical and the observed value of the observable quantity for each SNeIa, which is the distance modulus. The term $\mathbf{C}_{SN}$ gives the total covariance matrix.
Once a specific cosmological model has been chosen, the predicted distance modulus of the SNe, $\mu$, is defined as
\begin{equation}\label{eq:m_jla}
\mu(z,\boldsymbol{\theta}) = 5 \log_{10} [ d_{L}(z, \boldsymbol{\theta}) ] +\mu_0 \; ,
\end{equation}
where $D_{L}$ is the dimensionless luminosity distance given by
\be
d_L(z,\theta_c)=(1+z)\int_{0}^{z}\frac{dz'}{E(z')} \,,
\ee
with $E(z)=H(z)/H_0$. Notwithstandingly, our  $\chi^2_{SN}$ above would contain the nuisance parameter $\mu_0$, which in turn is a function  of the Hubble constant, the speed of light $c$ and the SNeIa absolute magnitude. This inconvenient degeneracy intrinsic to the definition of the parameters can be dealt with if we marginalize analytically over $\mu_0$. We refer the reader to \cite{conley} for details on this procedure.

The $\chi^2$ estimator thus obained reads
\be\label{eq:chis}
\chi^2_{SN}=a+\log \left(\frac{d}{2\pi}\right)-\frac{b^2}{d},
\ee
where $a\equiv\left(\Delta \boldsymbol{\mathcal{F}}_{SN}\right)^T \; \cdot \; \mathbf{C}^{-1}_{SN} \; \cdot \; \Delta  \boldsymbol{\mathcal{F}}_{SN}$, $b\equiv\left(\Delta \boldsymbol{\mathcal{F}}^{SN}\right)^T \; \cdot \; \mathbf{C}^{-1}_{SN} \; \cdot \; \boldsymbol{1}$ and $d\equiv\boldsymbol{1}\; \cdot \; \mathbf{C}^{-1}_{SN} \; \cdot \;\boldsymbol{1}$, with $\boldsymbol{1}$ being the identity matrix.

\subsection{Quasars}

The  distance modulus $\mu$ of some 808 quasars in the redshift range $0.06<z<6.28$ can be constructed from their UV and X-ray fluxes. (Following Eq.~(5) in \cite{Risaliti:2015zla}).

In order to test our model we are interested in the distance modulus $\mu$, so we compute it using Eq.~(5) given in \cite{Risaliti:2015zla}, i.e.,
\be
\mu=\frac{5}{2(\gamma-1)}[\log(F_X)-\gamma\log(F_{UV})-\beta']\,,
\ee
where $\gamma=0.6$ is the average value of the free parameter which relates both fluxes and $\beta'$ is a free scaling factor. As before we compute the theoretical distance modulus using Eq.~(\ref{eq:m_jla}) and marginalize over the additive constant terms so that the final $\chi^2$ is given by Eq.~(\ref{eq:chis})
where now we have: $a\equiv \left(\Delta\boldsymbol{\mathcal{F}}^{Q}\right)^T \, \cdot \, \mathbf{C}^{-1}_{Q} \, \cdot \, \Delta  \boldsymbol{\mathcal{F}}^{Q}$, $b\equiv\left(\Delta \boldsymbol{\mathcal{F}}^{Q}\right)^T \, \cdot \, \mathbf{C}^{-1}_{Q} \, \cdot \, \boldsymbol{1}$ and $d\equiv\boldsymbol{1}\, \cdot \, \mathbf{C}^{-1}_{Q} \, \cdot \, \boldsymbol{1}$.

\subsection{Gamma Ray Bursts}

We work with the Mayflower sample, which consist on 79  Gamma Ray Bursts (GRBs) and cover the redshift range $1.44<z<8.1$ \cite{Liu:2014vda}. We use  Eq.~(\ref{eq:m_jla}) to compute the theoretical distance modulus and then marginalize over the constant additive term. This way, the final $\chi^2$ estimator is given by Eq.~(\ref{eq:chis})
with $a\equiv \left(\Delta\boldsymbol{\mathcal{F}}^{G}\right)^T \, \cdot \, \mathbf{C}^{-1}_{G} \, \cdot \, \Delta  \boldsymbol{\mathcal{F}}^{G}$, $b\equiv\left(\Delta \boldsymbol{\mathcal{F}}^{G}\right)^T \, \cdot \, \mathbf{C}^{-1}_{G} \, \cdot \, \boldsymbol{1}$ and $d\equiv\boldsymbol{1}\, \cdot \, \mathbf{C}^{-1}_{G} \, \cdot \, \boldsymbol{1}$.

\subsection{Baryon Acoustic Oscillations}

In this section, we consider the $\chi^2_{BAO}$ estimator for Baryon Acoustic Oscillations (BAO), which is defined in the following manner
\begin{equation}
\chi^2_{BAO} = \Delta \boldsymbol{\mathcal{F}}^{BAO} \, \cdot \ \mathbf{C}^{-1}_{BAO} \, \cdot \, \Delta  \boldsymbol{\mathcal{F}}^{BAO} \ ,
\end{equation}
where the quantity $\mathcal{F}^{BAO}$ depends on the survey which is considered. We use data from the WiggleZ Dark Energy Survey, which are evaluated at redshifts $0.44$, $0.6$ and $0.73$, and are shown in Table~1 of \cite{Blake:2012pj}. The quantities we consider are the acoustic parameter
\begin{equation}\label{eq:AWiggle}
A(z) = 100  \sqrt{\Omega_{m} \, h^2} \frac{D_{V}(z)}{c \, z} \, ,
\end{equation}
and the Alcock-Paczynski distortion parameter, given by the following relation
\begin{equation}\label{eq:FWiggle}
F(z) = (1+z)  \frac{D_{A}(z)\, H(z)}{c} \, ,
\end{equation}
where $D_{A}$ is the angular diameter distance
\begin{equation}\label{eq:dA}
D_{A}(z)  = \frac{c}{H_{0}} \frac{1}{1+z} \ \int_{0}^{z} \frac{\mathrm{d}z'}{E(z')} \; ,
\end{equation}
and $D_{V}$ is the geometric mean of the physical angular diameter distance, $D_A$, and of the Hubble function, $H(z)$, and is written as
\begin{equation}\label{eq:dV}
D_{V}(z)  = \left[ (1+z)^2 D^{2}_{A}(z) \frac{c \, z}{H(z,\boldsymbol{\theta})}\right]^{1/3}.
\end{equation}
Moreover, we have considered the data from the SDSS-III Baryon Oscillation Spectroscopic Survey (BOSS) DR$12$, described in \cite{Alam:2016hwk} and given by
\begin{equation}
D_{M}(z) \frac{r^{fid}_{s}(z_{d})}{r_{s}(z_{d})}, \qquad H(z) \frac{r_{s}(z_{d})}{r^{fid}_{s}(z_{d})} \,.
\end{equation}
Here, the factor $r_{s}(z_{d})$ denotes the sound horizon evaluated at the dragging redshift $z_{d}$, the quantity $r^{fid}_{s}(z_{d})$ is the same sound horizon but calculated for a given fiducial cosmological model used, and is equal to $147.78$ Mpc \cite{Alam:2016hwk}.

We consider the approximation \cite{Eisenstein}
\begin{equation}\label{eq:zdrag}
z_{d} = \frac{1291 (\Omega_{m} \, h^2)^{0.251}}{1+0.659(\Omega_{m} \, h^2)^{0.828}} \left[ 1+ b_{1} (\Omega_{b} \, h^2)^{b2}\right]\; ,
\end{equation}
for the redshift of the drag epoch, where the factors $b_1$ and $b_2$ are given by
\begin{eqnarray}\label{eq:zdrag_b}
b_{1} &=& 0.313 (\Omega_{m} \, h^2)^{-0.419} \left[ 1+0.607 (\Omega_{m} \, h^2)^{0.6748}\right] \,,
	\nonumber \\
b_{2} &=& 0.238 (\Omega_{m} \, h^2)^{0.223}\,, \nonumber
\end{eqnarray}
respectively. In addition to this, the sound horizon is defined as:
\begin{equation}\label{eq:soundhor}
r_{s}(z) = \int^{\infty}_{z} \frac{c_{s}(z')}{H(z')} \mathrm{d}z'\, ,
\end{equation}
where the sound speed is given by
\begin{equation}\label{eq:soundspeed}
c_{s}(z) = \frac{c}{\sqrt{3(1+\overline{R}_{b}\, (1+z)^{-1})}} \; ,
\end{equation}
and $\overline{R}_{b}$ is defined as
\begin{equation}
\overline{R}_{b} = 31500 \Omega_{b} \, h^{2} \left( T_{CMB}/ 2.7 \right)^{-4}\,,
\end{equation}
with $T_{CMB} = 2.726$ K.
We have taken into account the point $D_V(z=1.52)=3843\pm 147\frac{r_s(zd)}{r_s^{fid}(z_d)}$ Mpc \cite{Ata:2017dya} from the  extended Baryon Oscillation Spectroscopic Survey (eBOSS).

We have also added data points from Quasar-Lyman $\alpha$ Forest from SDSS-III BOSS DR$11$ \cite{Font-Ribera:2013wce}:
\begin{eqnarray}
\frac{D_{A}(z=2.36)}{r_{s}(z_{d})} &=& 10.8 \pm 0.4\,, \\
\frac{c}{H(z=2.36) r_{s}(z_{d})}  &=& 9.0 \pm 0.3\,,
\end{eqnarray}
in our analysis.

\subsection{Cosmic Microwave Background data}

The $\chi^2_{CMB}$ estimator for the Cosmic Microwave Background (CMB) is defined as
\begin{equation}
\chi^2_{CMB} = \Delta \boldsymbol{\mathcal{F}}^{CMB} \; \cdot \; \mathbf{C}^{-1}_{CMB} \; \cdot \; \Delta  \boldsymbol{\mathcal{F}}^{CMB} \; ,
\end{equation}
where $\mathcal{F}^{CMB}$ is a vector of quantities used in \cite{PhysRevD.94.083521}, and we have considered the \textit{Planck} $2015$ data release in order to give the shift parameters defined in \cite{Wang:2007mza}. These parameters are related to the positions of the CMB acoustic peaks that depend on the geometry of the specific model considered and, as such, can be used to discriminate between dark energy models of the different nature. Their definitions are:
\begin{eqnarray}
R(\boldsymbol{\theta}) &\equiv& \sqrt{\Omega_m H^2_{0}} \frac{r(z_{\ast},\boldsymbol{\theta})}{c}, \nonumber \\
l_{a}(\boldsymbol{\theta}) &\equiv& \pi \frac{r(z_{\ast},\boldsymbol{\theta})}{r_{s}(z_{\ast},\boldsymbol{\theta})}\, .
\end{eqnarray}

As in the previous data sets, $r_{s}$ is the comoving sound horizon, evaluated at the photon-decoupling redshift $z_{\ast}$, given by the fitting formula \cite{Hu:1995en}:
\begin{eqnarray}{\label{eq:zdecoupl}}
z_{\ast} &=& 1048 \left[ 1 + 0.00124 (\Omega_{b} h^{2})^{-0.738}\right] \times   \nonumber \\
&& \times \left(1+g_{1} (\Omega_{m} h^{2})^{g_{2}} \right)  \,,
\end{eqnarray}
where the factors $g_1$ and $g_2$ are given by
\begin{eqnarray}
g_{1} &=& \frac{0.0783 (\Omega_{b} h^{2})^{-0.238}}{1+39.5(\Omega_{b} h^{2})^{-0.763}}\,, \nonumber \\
g_{2} &=& \frac{0.560}{1+21.1(\Omega_{b} h^{2})^{1.81}} \,, \nonumber
\end{eqnarray}
respectively, and $r$ is the comoving distance defined as
\begin{equation}
r(z, \boldsymbol{\theta} )  = \frac{c}{H_{0}} \int_{0}^{z} \frac{\mathrm{d}z'}{E(z',\boldsymbol{\theta})} \mathrm{d}z'\; .
\end{equation}

\subsection{Monte Carlo Markov Chain (MCMC)}

We have implemented an MCMC code \cite{Lazkoz:2010gz,Capozziello:2011tj} so that we can test the predictions of our theory with the available data. The goal of this technique is to minimize the total $\chi^2$ defined as
\begin{equation}
\chi^{2}= \chi_{H}^{2} + \chi_{SN}^{2}+ \chi_{Q}^{2}+ \chi_{G}^{2} + \chi_{BAO}^{2} + \chi_{CMB}^{2} \, .
\end{equation}

Also, in order to set up the degree of reliability of the models, we use the Bayesian Evidence, $\mathcal{E}$, which is in general  recognized as the most reliable statistical comparison tool, despite some issues related to the choice of the priors \cite{Nesseris:2012cq}, which we overcome by choosing wide-enough priors ranges. We compute it using the algorithm described in \cite{Mukherjee:2005wg}. After calculating the  Bayesian Evidence, one can obtain the Bayes Factor, which is defined as the ratio of evidences of two models, $M_{i}$ and $M_{j}$, $\mathcal{B}^{i}_{j} = \mathcal{E}_{i}/\mathcal{E}_{j}$. If $\mathcal{B}^{i}_{j} > 1$,  model $M_i$ is preferred over $M_j$, given the data. We have used the $\Lambda$CDM model, as the reference model $M_j$.

Even if the Bayes Factor is $\mathcal{B}^{i}_{j} > 1$, one cannot state how much better model $M_i$ is with respect to model $M_j$. For this purpose, we choose the so-called Jeffreys' Scale \cite{Jeffreys98}. Generally, Jeffreys' Scale affirms that: if $\ln \mathcal{B}^{i}_{j} < 1$, the evidence in favor of model $M_i$ is not significant; if $1 < \ln \mathcal{B}^{i}_{j} < 2.5$, the evidence is substantial; if $2.5 < \ln \mathcal{B}^{i}_{j} < 5$, it is strong; and if $\ln \mathcal{B}^{i}_{j} > 5$, it is decisive. Values of $\ln \mathcal{B}^{i}_{j}$ which are negative can be seen as evidence against model $M_i$ (or in favor of model $M_j$). In \cite{Nesseris:2012cq}, it is demonstrated that Jeffreys' scale is not a fully-reliable tool to compare models, but at the same time the statistical validity of the Bayes factor as an efficient model-comparison tool is not doubted: a Bayes factor $\mathcal{B}^{i}_{j}>1$ firmly states that the model $i$ is more likely than model $j$. In the following, we present our results in both contexts so that the  reader can make his/her interpretation.

\section{Results of the observational tests}\label{sec:results}

After implementing the MCMC code, we have obtained the values for the background parameters that appear in Table \ref{resultados1}. There are some general results that can be summarized before analyzing in detail model by model. It can be noticed that the parameters $\Omega_m$ and $\Omega_b$ do not present a relevant variation from one model to another, being in perfect agreement with \textit{Planck}'s predictions \cite{Aghanim:2018eyx} in every case. Some more interesting feature can be observed in the value of the Hubble constant, defined as $H_0=100~h$. With respect to the existing tension between the value obtained by local measurements \cite{Riess:2019cxk} and the one predicted by \textit{Planck}, our result for $H_0$ are perfectly consistent with the \textit{Planck} value.

\subsection{Model 1: $\Lambda$CDM}

The first model we have considered consists on a constant term plus a third and a fourth order term. These are exactly the terms which correspond to a $f(z)$ theory mimicking a $\Lambda$CDM background in the $Q$-formalism, i.e., Eq.($\ref{fl}$). According to this equation, one may expect that the values of $f_0,~f_3$ and $f_4$ are somehow related with $\Omega_m$ and $\Omega_r$ (an analog computation was done in  \cite{Lazkoz:2018aqk}). In fact, we could identity the quantities
\bea\label{11}
f_0&=&-6(1-\Omega_m-\Omega_r),\\
f_3&=&-6\Omega_m,\\\label{22}
f_4&=&-6\Omega_r.
\eea
If we substitute here the values of $\Omega_m,~\Omega_r$ for $\Lambda$CDM according to Table \ref{resultados1} we obtain the following numerical values for Eqs. (\ref{11})-(\ref{22}):
\be
f_0=-4.1, \qquad  f_3=-1.88, \qquad  f_4=-0.00053.
\ee
Comparing with the values of $f_0,~f_3,~f_4$ in the table for this model, we see that for the parameters $f_3$ anf $f_4$ there is an agreement at $1\sigma$ level. Nevertheless, we can also see that the parameter $f_0$ agrees in the order of magnitude but presents a discrepancy with respect to the one in the table.

{\renewcommand{\tabcolsep}{1.5mm}
{\renewcommand{\arraystretch}{1.5}
\begin{table*}[ht!]
\caption{Results}\label{resultados1}
\begin{minipage}{0.98\textwidth}
\centering
\resizebox*{\textwidth}{!}{
\begin{tabular}{c|ccccccccccc|cc|}
model & $\Omega_m $ &$\Omega_b$& $h$ &$f_0$&$f_{16}$ &$f_{14}$&$f_{12}$  &$f_{1}$&$f_{2}$  & $f_3$&$f_4$ &  $\mathcal{B}^{i}_{\Lambda}$ & $\ln \mathcal{B}^{i}_{\Lambda}$\\
\hline
1&$0.313_{-0.007}^{+0.007}$&$0.046_{-0.002}^{0.002}$&$0.69_{-0.02}^{+0.02}$&$-9.6_{-0.8}^{+0.7}$&-&-&-&-&-&$-1.95_{-0.04}^{+0.04}$&$-0.00046_{-0.00004}^{+0.00004}$&$1$&$0$\\
\hline
 3&$0.317_{-0.007}^{+0.006}$&$0.047_{-0.002}^{+0.003}$&$0.69_{-0.02}^{+0.02}$&-&-&-&$-22.1_{-2.2}^{+1.7}$&$8.2_{-3.3}^{+2.2}$&$-0.5_{-0.5}^{+0.3}$&$-2.06_{-0.10}^{+0.09}$&$-0.00039_{-0.00005}^{+0.00004}$ &$0.394$  &  $-0.931$\\
 \hline
4&$0.317_{-0.007}^{+0.007}$&$0.047_{-0.002}^{+0.002}$&$0.69_{-0.02}^{+0.02}$&-&-&$-14.3_{-1.4}^{+1.3}$&-&-&-&$-2.06_{-0.04}^{+0.05}$&$-0.00040_{-0.00004}^{+0.00004}$ &$0.712$    &$-0.339$ \\
\hline
5&$0.315_{-0.006}^{+0.007}$&$0.047_{-0.003}^{+0.003}$&$0.69_{-0.02}^{+0.02}$&-&-&$-13.0_{-2.1}^{+1.6}$&-&$2.2_{-1.0}^{+0.9}$&$-0.28_{-0.27}^{0.46}$&$-1.99_{-0.05}^{+0.06}$&$-0.00045_{-0.00004}^{+0.00005}$&$0.791$    &$-0.234$\\
\hline
6&$0.314_{-0.008}^{+0.009}$&$0.046_{-0.002}^{+0.003}$&$0.69_{-0.02}^{+0.02}$&-&$-12.5_{-1.4}^{+1.2}$&-&-&-&-&$-2.06_{-0.04}^{+0.05}$&$-0.00042_{-0.00004}^{+0.00004}$ &$0.917$ &  $-0.087$
 \\
\hline
7&$0.315_{-0.007}^{+0.008}$&$0.047_{-0.002}^{+0.003}$&$0.69_{-0.02}^{+0.02}$&-&$-12.2_{-1.0}^{+1.2}$&-&-&$0.05_ {-0.07}^{+0.14}$&$0.004_{-0.053}^{+0.037}$&$-2.01_{-0.04}^{+0.04}$&$-0.00042_{-0.00004}^{+0.00003}$&$1.011 $    &$0.011$
\end{tabular}}
\label{resultados}
\end{minipage}
\end{table*}}}

\subsection{Models 2-3}

These two models do not include a constant term which could play the role of an effective cosmological constant. Instead, they are characterized by including a $1/2$ power in the polynomial; and model 3 also presents an order one and an order two term, corresponding to the parameteres $f_1$ and $f_2$. It must be noticed that model 2 does not appear in the table as it does not reach a $\chi^2$ which stabilizes and, moreover, it is higher than all other cases, which means it is clearly discarded by data. For model 3 it is not possible to make a concrete interpretation of the values of $f_1$ and $f_2$, because the corresponding histograms are highly irregular and far from being gaussian, so that the error bars are just an approximate and indicative estimation of their parameter space width. However, there is one interesting manner we can extract information from these two quantities: when looking at the parameters $f_3$ and $f_4$, we see their values are not very far (and statistically consistent) with those from model 1. Thus, we can infer that the role played by $f_1$ and $f_2$ is marginal. Moreover, the Bayes Factor of model $3$ is the highest in the sample; which means that it is the most disfavoured by data.

\subsection{Models 4-7}

As well as models 2 and 3, models 4-7 do not include a constant term, but they perform much better. Model 4 consists on a $1/4$ polynomial term plus a third and a fourth order term. Model 5 includes the same free parameters plus the $f_1$ and $f_2$ terms. The value of $f_{14}$ does not vary significantly between these two models, presenting an agreement at $1\sigma$ level. On the one hand, it is interesting to notice that the parameter $f_1$ is consistent with zero at approximately $2\sigma$, while $f_2$ contains zero at $1\sigma$. Thinking about results for model $3$, these results could confirm that the presence of any effect of such order on cosmological scales can be quite confidently discarded. Finally, for $f_3$ and $f_4$, the obtained values are quite in concordance with those of model 1.

Model 6 and 7 have been constructed in a similar way. Model 6 is described by a $1/6$ polynomial term plus the usual third and fourth order ones, while model 7 contains, in addition, $f_1$ and $f_2$ parameters. Again, the quantity corresponding to the smallest power, $f_{16}$ in this case, does not vary a lot between model 6 and 7.  Moreover, in this case we have found more significant results for $f_1$ and $f_2$, as they are both consistent with zero at $1\sigma$ level.  Again, $f_3$ and $f_4$ seem not to depend much on the model, having similar values in every model. Thus, we can infer from these results that cosmological data shows a statistical preference for small-order terms, i.e. for a lower-order dependence of the additional terms with redshift. In other words, even in the $f(Q)$ approach, data seem to confirm that any alternative solution must behave closely as a cosmological constant, at least for what concerns the cosmological background.

With respect to the reliability of the models in comparison with the $\Lambda$CDM model, the differences in the Bayes factor and Jeffreys' scale are so small that  we cannot extract much information from them. Although one cannot set decisive conclusions about the preference of one model with respect to the standard $\Lambda$CDM, this new formalism seems to be a viable alternative within the study of the late-time expansion of the universe.

\section{Conclusions}\label{sec:conclusions}

Within the context of modified gravity we have chosen a novel geometrical scenario based on the nonmetricity, $Q$, to obtain observational constraints for the background quantities of the late-time universe. We have selected the $f(Q)$ type of theories and followed the $f(z)$ approach to give some phenomenologically motivated models to test against observational data. We have used a wide variety of observational data sets to check the validity of our proposals. As an interesting result we have seen that order one and order two terms, i.e., $f_1$ and $f_2$, are, in general, compatible with zero. Moreover, the values we obtain for the rest of parameters are very similar for a model with and without these two parameters, not showing a significant dependence on their presence. This result coincides with the one obtained in the paper whose approach we are following.

We have also found that some of the modified parameters that we introduce, $f_3$ and $f_4$ more specifically, could be related in some way with the background parameters $\Omega_m$ and $\Omega_r$. Finally, an interesting result we want to remark is that the value obtained for the Hubble parameter, $H_0$, lies close to the \textit{Planck} estimation.

From the statistical point of view we cannot state a clear preference with 
respect to any of the considered models. In general, the values of the Bayes factor and their 
interpretation in terms of the Jeffreys' scale show a marginal and weak evidence in favor of the 
scenarios which slightly depart from the standard $\Lambda$CDM model. Nevertheless, 
one should not forget that that $f(R)/f(Q)$ scenarios have an inherent 
multidimensionality as regards the space of parameters, superior to that of standard dark 
energy settings. In spite of that, we have not fine-tuned our extended models with any a priori 
constraint, but we have left the corresponding parameters totally free to vary. Thus, the post-analysis consistency of the values of the background parameters for such models is 
a further asset (which somehow escapes the pure statistical evaluation) and can be considered 
as a hint pointing to continue the study of alternative theories of gravity within this kind of 
approach. This work can be understood as a first step in the observational study of a new class 
of modified gravity theories which lay in a geometry described by the nonmetricity. By going deeper in the cosmological and observational studies of these yet unexplored theories we may find interesting information which may contribute to our understanding of the evolution of the late-time universe.

\section*{Acknowledgments}
RL and MOB were supported by the Spanish Ministry of Economy and Competitiveness through research projects No. FIS2014-57956-P (comprising FEDER funds) and also by the Basque Government through research
project No. GIC17/116-IT956-16. FSNL acknowledges funding from the research grants No. UID/FIS/04434/2019, No. PTDC/FIS-OUT/29048/2017 and CEECIND/04057/2017.
MOB acknowledges financial support from the FPI grant BES-2015-071489. This article is based upon work from COST Action CA15117 (CANTATA), supported by COST (European Cooperation in Science and Technology).

\bibliography{biblio}{}

\end{document}